\documentclass{aa} 
\usepackage{graphicx}
\usepackage{txfonts}
\usepackage{natbib}
\bibpunct{(}{)}{;}{a}{}{,}

\begin{document}

   \title{Steps toward a high precision solar rotation profile:
   Results from SDO/AIA coronal bright point data}

   \author{D. Sudar
          \inst{1}
          \and
          I. Skoki\'{c}
          \inst{2}
          \and
          R. Braj\v{s}a
          \inst{1}
          \and
          S.~H. Saar
          \inst{3}
          }

   \institute{Hvar Observatory, Faculty of Geodesy, University of Zagreb,
              Ka\v{c}i\'{c}eva 26, 10000 Zagreb, Croatia
              \and
              Cybrotech Ltd, Bohinjska 11, 10000 Zagreb, Croatia
              \and
              Harvard-Smithsonian Center for Astrophysics, 60 Garden Street, Cambridge, MA 02138, USA
             }

\offprints{D. Sudar, \email: davor.sudar@gmail.com}

   \date{Release \today}
\abstract {Coronal bright points (CBP) are ubiquitous small brightenings in the solar corona associated with small magnetic bipoles.}
{We derive the solar differential rotation profile by tracing the motions of
CBPs detected by the Atmospheric Imaging Assembly (AIA)
instrument aboard the Solar Dynamics
Observatory (SDO). We also investigate
problems related to detection of coronal bright points resulting from instrument
and detection algorithm limitations.}
{To determine the positions and identification of coronal bright points we used
a segmentation algorithm. A linear fit of their central meridian distance and
latitude versus time was utilised to derive velocities.}
{We obtained 906
velocity measurements in a time interval
of only 2 days. The differential rotation profile can be expressed as
$\omega_{rot} =  (14.47\pm 0.10  + (0.6\pm 1.0)\sin^{2}(b) + (-4.7\pm 1.7)\sin^{4}(b))$\degr day$^{-1}$. Our result
is in agreement with other work and it comes with reasonable errors in spite of the very short time interval used.
This was made possible by the higher sensitivity and resolution of the AIA 
instrument compared to similar equipment as well as high cadence. The segmentation algorithm also played a crucial role by detecting
so many CBPs, which reduced the errors to a reasonable level.}
{Data and methods presented in this paper show a great potential to obtain very accurate
velocity profiles, both for rotation and meridional motion and, consequently,
Reynolds stresses. The amount of coronal bright point data that could be obtained from this
instrument should also provide a great opportunity to study changes of velocity
patterns with a temporal resolution of only a few months. Other possibilities are studies of evolution
of CBPs and proper motions of magnetic elements on the Sun.}
\keywords{Sun: rotation - Sun: corona - Sun: activity}
   \maketitle

\section{Introduction}
We present a new solar rotation profile obtained by tracing the motions of
coronal bright points (CBPs) observed by Atmospheric Imaging Assembly
(AIA) instrument on board the Solar Dynamics Observatory (SDO) satellite \citep{Lemen2012}.

The most frequently used and oldest tracers of  the solar differential rotation profile
are sunspots \citep{Newton1951, Howard1984, Balthasar1986, Brajsa2002}.
One of the advantages of using sunspots is very long time coverage.
On the other hand, there are numerous disadvantages: sunspots have
complex and evolving structure, their distribution in latitude
is highly non-uniform and it does not extend to
higher solar latitudes. The number of sunspots is also highly
variable during the solar cycle which makes measurements of
solar differential rotation profile almost impossible during solar minimum.

CBPs are more uniformly distributed in latitude
and are numerous in all phases of the solar cycle. They also extend
over all solar latitudes. They have been used as tracers of solar rotation
since the beginning of the space age \citep{Dupree1972}.
In recent years there are numerous studies investigating solar
differential rotation by using CBPs as tracers. \citet{Kariyappa2008, Hara2009}
used Yohkoh/SXT data while \citet{Brajsa2001, Brajsa2002b, Brajsa2004, Vrsnak2003, Wohl2010} used
SOHO-EIT observations in 28.4 nm channel and \citet{Karachik2006} used 19.4 nm SOHO/EIT channel.
\citet{Kariyappa2008} also used Hinode/XRT
full-disk images to determine the solar rotation profile.

Other tracers are used as well: magnetic fields \citep{Wilcox1970, Snodgrass1983,
Komm1993} and H$\alpha$ filaments \citep{Brajsa1991}. Apart from tracers, Doppler measurements
can also be used \citep{Howard1970, Ulrich1988, Snodgrass1990}.

Helioseismic measurements also show differential rotation below the photosphere
all the way down to the bottom of the convective zone \citep{Kosovichev1997,
Schou1998}. Further down, the rotation profile becomes uniform for all
latitudes \citep[cf. eg.][]{Howe2009}.

For further details about solar rotation, its importance for solar dynamo models, and comparison of rotation measurements
between different sources, see the reviews by \citet{Schroeter1985, Howard1984rev, Beck2000, Ossendrijver2003,
Rudiger2004, Stix2004, Howe2009, Rozelot2009}.

In this work we use CBP data obtained by SDO/AIA over only two days to assess the quality
of the data, identify sources of errors and calculate the solar differential rotation profile.
We will also investigate the possibility of using CBP data from SDO/AIA for further
studies of other related phenomena (meridional flow, rotation velocity residuals
and Reynolds stress).

CBP data from SDO/AIA were also used in other works. \citet{Lorenc2012} discussed rotation
of the solar corona based on 69 structures from 674 images detected in 9.4 nm channel using an interactive method
of detection. \citet{Dorotovic2014} presented a hybrid algorithm for detection and tracking
of CBPs.
\citet{McIntosh2014a} used detection algorithm presented in their previous
paper \citep{McIntosh2005} to identify CBPs in SDO/AIA 19.4 nm channel and correlate their
properties with those of giant convective cells.
Using more SDO/AIA data and extending analysis back to SOHO era,
\citet{McIntosh2014b} concluded that CBPs almost exclusively form around the vertices
of giant convective cells.

\section{Data and reduction methods}
We have used data from the AIA instrument on board the SDO
satellite \citep{Lemen2012}.
The spatial resolution of the instrument is $\approx$0.6"/pixel. For comparison, SOHO/EIT resolution
is 2.629"/pixel while Hinode/XRT has a resolution of 1.032"/pixel.

To obtain positional information for the coronal bright points (CBPs),
we employed a segmentation algorithm which uses the 19.3 nm AIA channel data
to search for localized, small intensity enhancements in the EUV
compared to a smoothed background intensity. More details about the detection
algorithm, which is similar to the algorithm by \citet{McIntosh2005}, can
be found in \citet{Martens2012}.

This resulted in measurements of 66842 positions of 13646 individual
CBPs covering two days (1st and 2nd of January 2011).
The time interval between two successive images was 10 minutes.
\begin{figure}
\resizebox{\hsize}{!}{\includegraphics{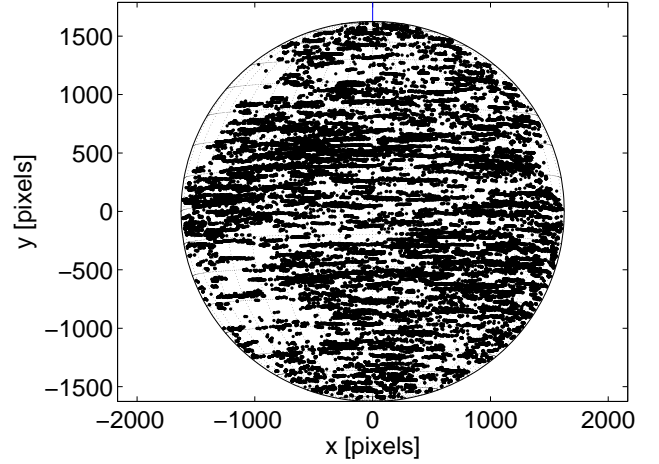}}
\resizebox{\hsize}{!}{\includegraphics{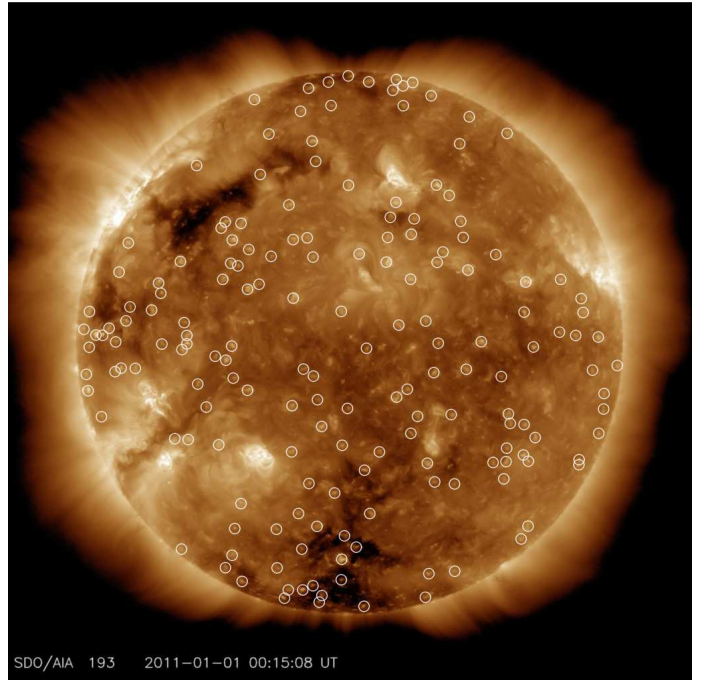}}
\caption{Distribution of CBPs detected by the segmentation algorithm (top panel)
and image of the Sun in the 19.3 nm channel obtained by SDO/AIA on 1st of January
2011. White circles show detected CBPs on this image (bottom panel).}
\label{Fig_CBP_distr_full}
\end{figure}
In top panel of Fig.~\ref{Fig_CBP_distr_full} we show the distribution of detected
CBPs and compare it to the full disk image of the Sun in the 19.3 nm channel obtained
on 1st of January 2011 (bottom panel of the same Figure).
In the bottom panel, white circles show CBPs that were detected on one image
by the segmentation algorithm.
We can see that CBPs are scarce in active regions, partly because of difficulties
in detecting them against such bright and variable backgrounds.

The segmentation algorithm provides coordinates in pixels (centroids of CBPs on the image)
and we converted them to heliographic coordinates taking into account the
current solar distance given in FITS files \citep{Rosa1995, Rosa1998}.
Positions of objects near the solar limb are fairly inaccurate. Limiting the data
to $\pm$58\degr from the centre of the Sun or $\approx$0.85$R_{\odot}$ of the projected solar disk
removes this problem \citep[cf.][]{Stark1981, Balthasar1986b}.

\begin{figure}
\resizebox{\hsize}{!}{\includegraphics{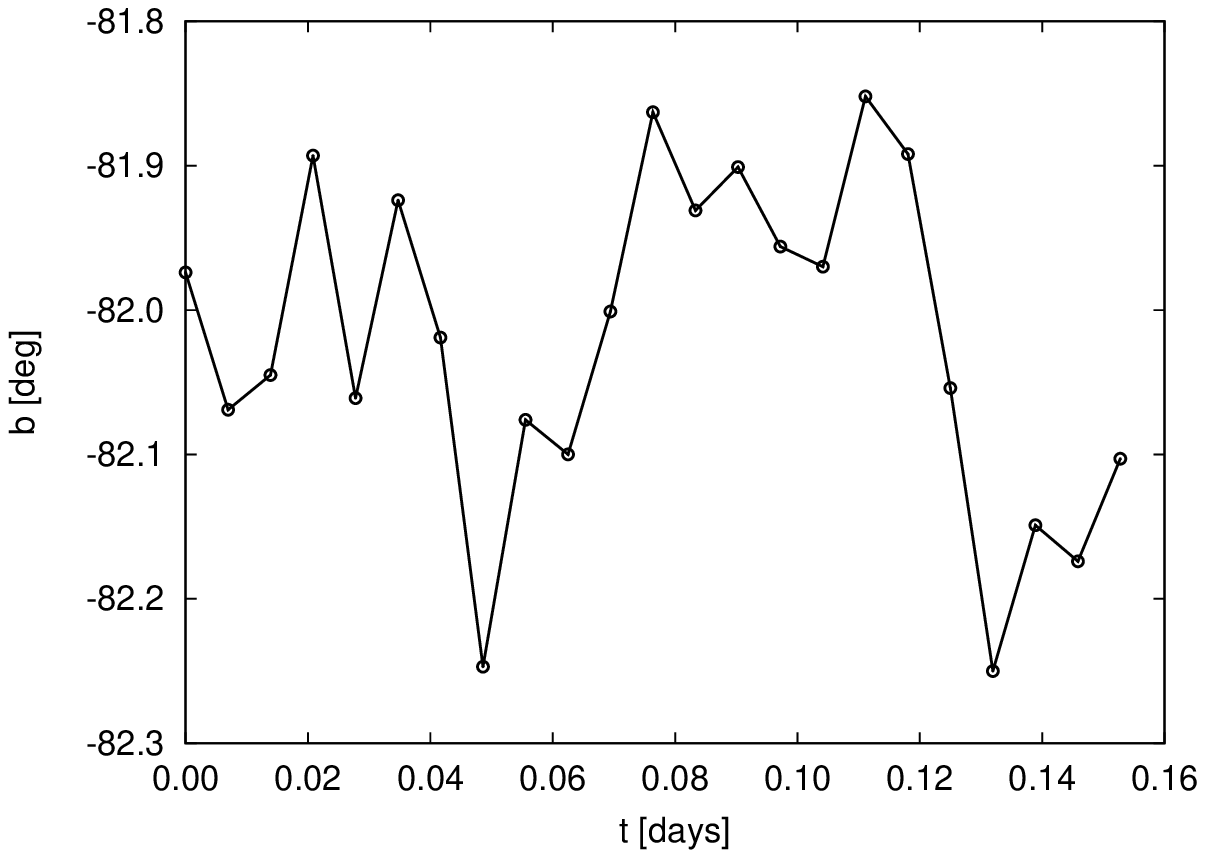}}
\resizebox{\hsize}{!}{\includegraphics{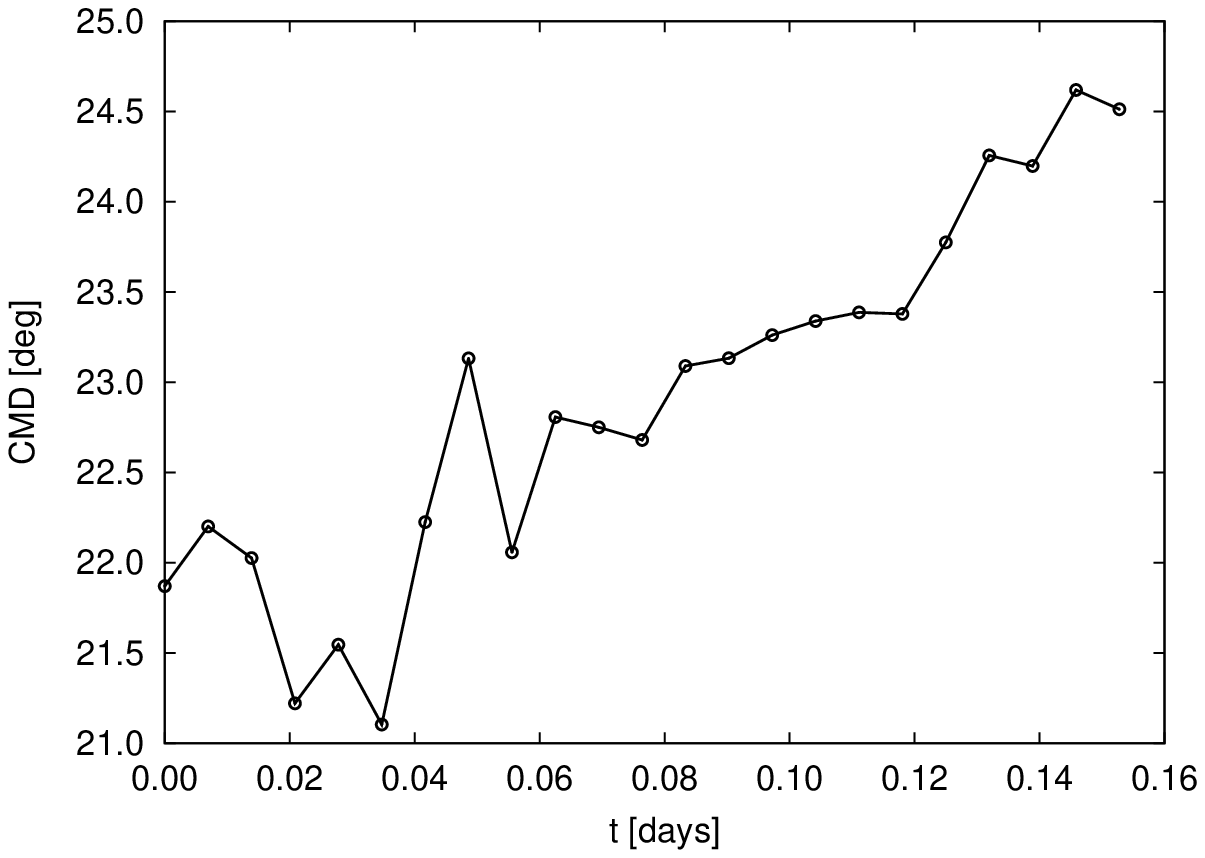}}
\caption{Motion of a single CBP. Top panel: latitude, $b$, over
time, $t$; bottom panel: central meridional distance, $CMD$ over time, $t$.}
\label{Fig_cbp_single_motion}
\end{figure}
As can be seen from Fig.~\ref{Fig_cbp_single_motion}, the calculated velocities
show some scatter.  This scatter arises because the shifts are fairly small at
our 10 min cadence, and there can be 
significant variations in brightness and structure of CBPs which 
influences calculation of the centroid points. Nevertheless, trends are visible,
especially in azimuthal motion which is known to be a significantly larger
effect.  The dominant azimuthal motion led us to approximate the CBP motion 
with a linear fit to calculate the velocities:
\begin{equation}
\omega_{syn}= \frac{N\sum\limits^{N}_{i=1} l_{i}t_{i} - \sum\limits^{N}_{i=1} l_{i}\sum\limits^{N}_{i=1} t_{i}}{N\sum\limits^{N}_{i=1} t^{2}_{i}-\left(\sum\limits^{N}_{i=1} t_{i}\right)^{2}},
\end{equation}
\begin{equation}
\omega_{mer}= \frac{N\sum\limits^{N}_{i=1} b_{i}t_{i} - \sum\limits^{N}_{i=1} b_{i}\sum\limits^{N}_{i=1} t_{i}}{N\sum\limits^{N}_{i=1} t^{2}_{i}-\left(\sum\limits^{N}_{i=1} t_{i}\right)^{2}},
\end{equation}
where $\omega_{syn}$ is a synodic rotational velocity, $\omega_{mer}$ is a meridional angular velocity,
$l_{i}$ is central meridian distance (CMD) and $b_{i}$ is latitude of each measurement for a single CBP.
We have also removed all CBPs which had less than 10 measurements of position
in order for linear fits to be more robust. This is equivalent to 100 minutes
or about 1\degr at the equator.
To obtain the true rotation of CBPs on the Sun we convert
synodic velocities to sidereal using Eq.~7 from \citet{Skokic2014}. 

Trying to identify the same object on subsequent images with an automatic
method is bound to result in some misidentification.
The resulting velocities are usually
very large and can easily be removed by applying a simple velocity filter.
Even the human factor can introduce
such errors. For example, \citet{Sudar2014} analysed solar rotation residuals
and meridional motions of sunspot groups from the Greenwich Photoheliographic Results
and found that they had to use a filter 8$<\omega_{rot}<$19\degr day$^{-1}$ for rotational
velocity in order to eliminate these erroneous measurements. The Greenwich Photoheliographic Results
catalogue is being investigated and revised partly in order to remove
such problems \citep{Willis2013a, Willis2013b, Erwin2013}.
In this work we have also used a 8$<\omega_{rot}<$19\degr day$^{-1}$ filter for rotational velocities
to remove such outliers. In addition, we applied a meridional velocity filter 
of -4$<\omega_{mer}<$4\degr day$^{-1}$ to remove further outliers.

After completing all the procedures described above, we obtained 906 velocity
measurements by tracing CBPs over just two days.
\citet{Olemskoy2005} pointed out that non-uniform distribution of tracers can result
in false flows. This effect is most notable for meridional motion and rotation velocity
residuals, but can easily be removed by assigning the calculated velocity to the 
latitude of the first measurement of position \citep{Olemskoy2005, Sudar2014}.
Although the effect is negligible for solar rotation, we nevertheless applied 
the correction in this work.

It is important to keep in mind that even when the tracers are 
uniformly distributed over the solar surface, the distribution
of tracers in latitude will be non-uniform. As we move from equator to the pole, the area of each latitude bin becomes smaller, so we observe
progressively fewer tracers ($\sim \cos{b}$).

\section{Results}
In this work, we present an analysis of the
motion of CBPs observed by the SDO/AIA instrument.
For a better understanding of the results and the potential of
future studies along these lines, it is very useful to analyse
the accuracy and errors of the dataset.

\begin{figure}
\resizebox{\hsize}{!}{\includegraphics{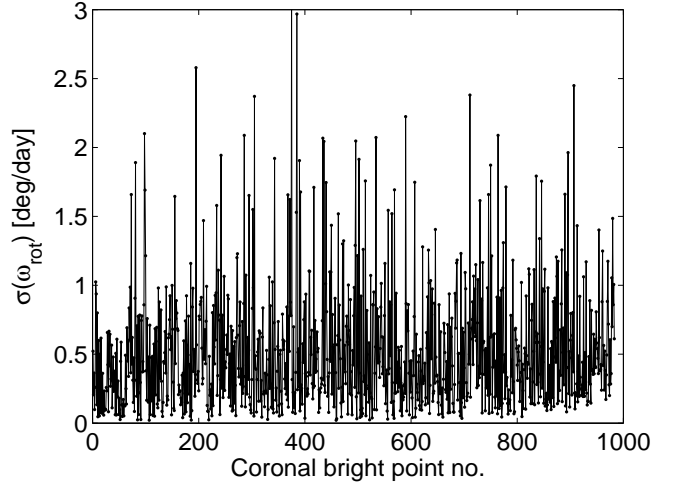}}
\caption{Errors of rotational velocity, $\sigma(\omega_{rot})$, for each of
the 906 measurements.}
\label{Fig_rot_errors_ind}
\end{figure}
\begin{figure}
\resizebox{\hsize}{!}{\includegraphics{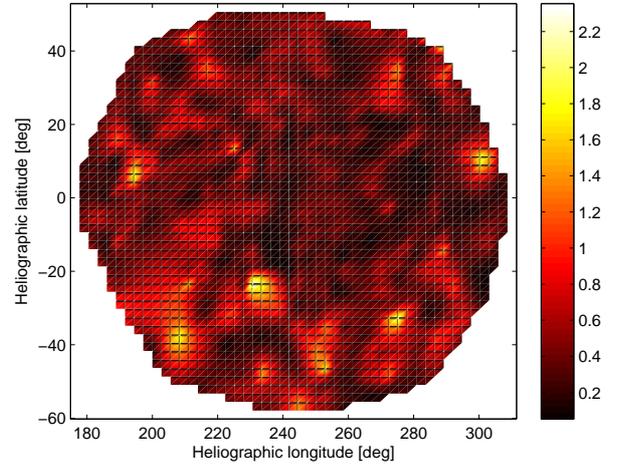}}
\caption{Distribution of rotation velocity errors ($\omega_{rot}$) in
heliographic coordinates. Error scale is in \degr\ day$^{-1}$.}
\label{Fig_rot_errors_lat_lon}
\end{figure}

In Fig.~\ref{Fig_rot_errors_ind} we show errors of the calculated
rotational velocities, $\sigma(\omega_{rot})$, for each CBP, which resulted
from errors in the linear fitting of longitude vs time measurements.
Although the
errors can go up to 3\degr day$^{-1}$, the majority is below
1\degr day$^{-1}$. In Fig.~\ref{Fig_rot_errors_lat_lon} we show these
errors in heliographic coordinates to check their spatial distribution
on the solar surface.
Larger errors are shown with brighter shades and we can see that
these roughly correspond to the positions of active regions shown in the bottom
panel of Fig.~\ref{Fig_CBP_distr_full}. This correlation with active
region is probably a consequence of the detection algorithm design and
difficulties in detection of CBPs over a bright, variable background.

\begin{figure}
\resizebox{\hsize}{!}{\includegraphics{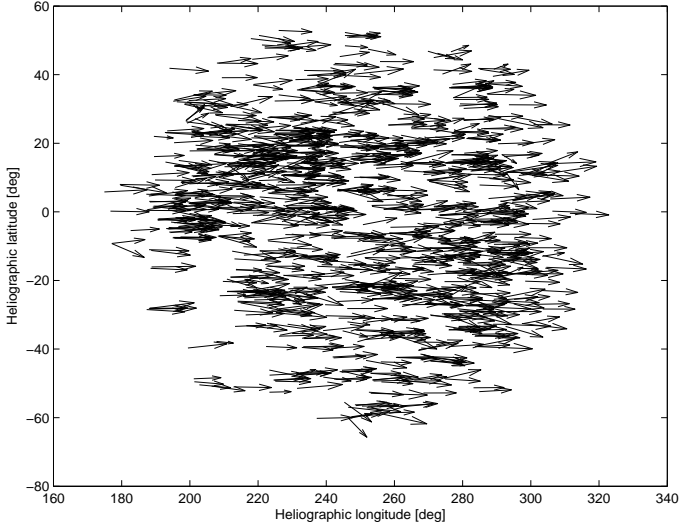}}
\caption{Distribution of CBPs in heliographic
coordinates with arrows showing the direction and strength of
the velocity vector.}
\label{Fig_rotation_vector_pos}
\end{figure}

In Fig.~\ref{Fig_rotation_vector_pos} we show the distribution
of CBPs in heliographic coordinates with arrows indicating the velocity vector.
As expected, the dominant effect is that of solar rotation.

The latitudinal dependence of rotational velocity is usually expressed
as \citep{Howard1970, Schroeter1985}:
\begin{equation}
\omega_{rot}(b)=A + B\sin^2{b} + C\sin^{4}{b},
\label{Eq_difrot_profile}
\end{equation}
where $b$ is the latitude. Parameter $A$ represents equatorial
velocity, while $B$ and $C$ depict the deviation from rigid body
rotation. The problem with Eq.~\ref{Eq_difrot_profile} is that the
functions in this expression are not orthogonal, so the parameters
are not independent of each other \citep{Duvall1978, Snodgrass1984,
Snodgrass1985, Snodgrass1990}. This crosstalk among the coefficients
is particularity bad for $B$ and $C$. The effect of crosstalk
does not affect the actual shape of the fit ($\omega_{rot}(b)$),
but it creates confusion when directly comparing coefficients
from different authors or obtained by different indicators.

There are various methods to alleviate this problem.  
Frequently $C$ is set to
zero since its effect is noticeable
only at higher latitudes. This is almost a standard practise when observing
rotation by tracing sunspots or sunspot groups because their
positions do not extend to high latitudes \citep{Howard1984,Balthasar1986,
Pulkkinen1998,Brajsa2002,Sudar2014}.

Another method to reduce the crosstalk problem is to set the $C/B$ ratio
to some fixed value. \citet{Scherrer1980} set the ratio $C/B = 1$
while \citet{Ulrich1988}, after measuring the covariance
of $B$ and $C$, set the ratio to $C/B=1.0216295$.

\begin{figure}
\resizebox{\hsize}{!}{\includegraphics{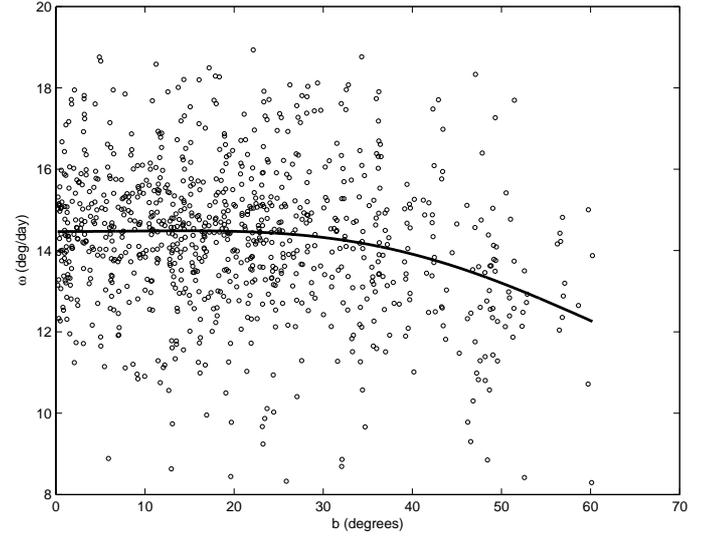}}
\caption{Solar differential rotation profile obtained with data from SDO/AIA.
Open circles are individual measurements, while the solid line is the
best fit defined by Eq.~\ref{Eq_difrot_profile} for the $A\neq B\neq C$
case.}
\label{Fig_dif_rot_profile_fit}
\end{figure}
In Fig.~\ref{Fig_dif_rot_profile_fit} we show individual measurements
of rotational velocities, $\omega_{rot}$, with respect to latitude, $b$,
as open circles. We indicate, with a solid line, the best
fit to the data using a functional form given in Eq.~\ref{Eq_difrot_profile}.
Coefficients of the fit are given in Table~\ref{Tab_coef_fit}.
We also fitted the rotation profile for Northern and Southern solar hemisphere
separately because of possible asymmetry \citep[cf.eg.][]{Wohl2010} and show
the results in the same table.
Coefficient $A$ shows a larger value in the Southern hemisphere for all 3 fit
functions. \citet{Jurdana2011} reported that coefficient $A$ is larger when
solar activity is smaller. According to the SIDC data \citep{sidc} we can see
that Northern hemisphere is more active both when looking at the monthly smoothed means
and daily sunspot data, consistent with \citet{Jurdana2011}.
However, judging by the errors of the coefficients, the difference between South
and North is statistically
low and the hypothesis that this is a result of asymmetric solar activity needs
to be verified with a larger data sample.
\begin{table}
\caption{Coefficients of the solar rotation profile.}
\label{Tab_coef_fit}
\centering
\begin{tabular}{c l l l c}
\hline\hline
Type & $A$ [\degr day$^{-1}$] & $B$ [\degr day$^{-1}$] & $C$ [\degr day$^{-1}$] & $n$\\
\hline
$A$, $B$, $C$ & 14.47$\pm$0.10 & +0.6$\pm$1.0 & -4.7$\pm$1.7 & 906\\
$A$, $B=C$ & 14.59$\pm$0.07 & -1.35$\pm$0.21 & -1.35$\pm$0.21 & 906\\
$A$, $B$, $C=0$ & 14.62$\pm$0.08 & -2.02$\pm$0.33 & 0 & 906\\
\hline
\multicolumn{5}{c}{Northern hemisphere}\\
\hline
$A$, $B$, $C$ & 14.43$\pm$0.13 & +0.8$\pm$1.5 & -5.6$\pm$3.0 & 461\\
$A$, $B=C$ & 14.55$\pm$0.10 & -1.35$\pm$0.35 & -1.35$\pm$0.35 & 461\\
$A$, $B$, $C=0$ & 14.57$\pm$0.10 & -1.92$\pm$0.52 & 0 & 461\\
\hline
\multicolumn{5}{c}{Southern hemisphere}\\
\hline
$A$, $B$, $C$ & 14.50$\pm$0.15 & +0.7$\pm$1.4 & -4.8$\pm$2.3 & 445\\
$A$, $B=C$ & 14.65$\pm$0.11 & -1.39$\pm$0.28 & -1.39$\pm$0.28 & 445\\
$A$, $B$, $C=0$ & 14.69$\pm$0.12 & -2.14$\pm$0.45 & 0 & 445\\
\hline
\end{tabular}
\end{table}

\begin{figure}
\resizebox{\hsize}{!}{\includegraphics{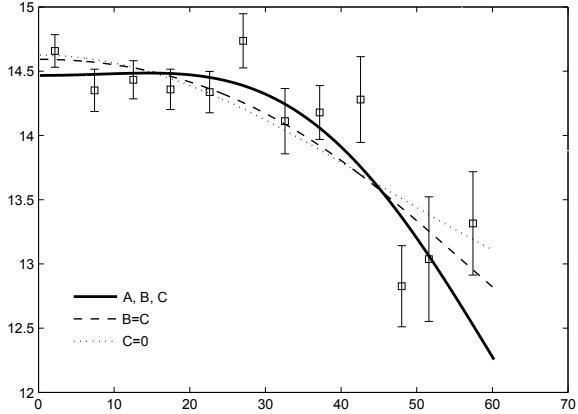}}
\caption{Comparison of three different fitting procedures for the solar differential
rotation profile. Average values of $\omega_{rot}$ in 5\degr\ bins in latitude, $b$,
are also shown with their respective errors.}
\label{Fig_dif_rot_profile_comp}
\end{figure}

In Fig.~\ref{Fig_dif_rot_profile_comp} we show a comparison between different
fitting techniques: $A\neq B\neq C$ (solid line), $A\neq B = C$ (dashed line)
and $A\neq B$, $C=0$ (dotted line). In the same figure we also show
average values of $\omega_{rot}$ in bins 5\degr\ wide in latitude, $b$,
with their respective errors.

\section{Discussion}
\begin{table*}
\caption{Comparison with some other results.}
\label{Tab_comp_fit}
\centering
\begin{tabular}{c c l l l l l l c}
\hline\hline
Method/object & time period & $A$ [\degr day$^{-1}$] & $B$ [\degr day$^{-1}$] & $C$ [\degr day$^{-1}$] & $A_{G}$ [\degr day$^{-1}$] & $B_{G}$ [\degr day$^{-1}$] & $C_{G}$ [\degr day$^{-1}$] & Ref\\
\hline
CBPs & 1967 & 14.65$\pm$0.2 & & & 14.65 & & & (1)\\
CBPs & 1994-1997 & 14.39$\pm$0.01 & -1.91$\pm$0.10 & -2.45$\pm$0.17 & 13.80 & -0.709 & -0.117 & (2)\\
CBPs & 1998-1999 & 14.454$\pm$0.027 & -2.22$\pm$0.07 & -2.22$\pm$0.07 & 13.82 & -0.740 & -0.106 & (3)\\
CBPs & 1998-2006 & 14.499$\pm$0.006 & -2.54$\pm$0.06 & -0.77$\pm$0.09 & 13.93 & -0.611 & -0.037 & (4)\\
CBPs & 1-2 Jan 2011 & 14.47$\pm$0.10 & +0.6$\pm$1.0 & -4.7$\pm$1.7 & 14.19 & -0.507 & -0.224 & (5)\\
CBPs & 1-2 Jan 2011 & 14.59$\pm$0.07 & -1.35$\pm$0.21 & -1.35$\pm$0.21 & 14.20 & -0.450 & -0.064 & (5)\\
CBPs & 1-2 Jan 2011 & 14.62$\pm$0.08 & -2.02$\pm$0.33 & & 14.22 & -0.404 & & (5)\\
\hline
sunspot groups & 1853-1996 & 14.531$\pm$0.003 & -2.75$\pm$0.05 & & 13.98 & -0.550 & & (6)\\
sunspot groups & 1874-1976 & 14.551$\pm$0.006 & -2.87$\pm$0.06 & & 13.98 & -0.574 & & (7)\\
sunspot groups & 1878-2011 & 14.499$\pm$0.005 & -2.64$\pm$0.05 & & 13.97 & -0.528 & & (8)\\
sunspot groups & 1880-1976 & 14.37$\pm$0.01 & -2.59$\pm$0.16 & & 13.85 & -0.518 & & (9)\\
sunspots & 1921-1982 & 14.522$\pm$0.004 & -2.84$\pm$0.04 & & 13.95 & -0.568 & & (10)\\
sunspot groups & 1921-1982 & 14.393$\pm$0.010 & -2.95$\pm$0.09 & & 13.80 & -0.590 & & (10)\\
\hline
H$\alpha$ filaments& 1972-1987 & 14.45$\pm$0.15 & -0.11$\pm$0.90 & -3.69$\pm$0.90 & 14.11 & -0.514 & -0.176 & (11)\\
\hline
magnetic features & 1967-1980 & 14.307$\pm$0.005 & -1.98$\pm$0.06 & -2.15$\pm$0.11 & 13.73 & -0.683 & -0.102 & (12)\\
magnetic features & 1975-1991 & 14.42$\pm$0.02 & -2.00$\pm$0.13 & -2.09$\pm$0.15 & 13.84 & -0.679 & -0.100 & (13)\\
\hline
Doppler & 1966-1968 & 13.76 & -1.74 & -2.19 & 13.22 & -0.640 & -0.104 & (14)\\
Doppler & 1967-1984 & 14.05 & -1.49 & -2.61 & 13.53 & -0.646 & -0.124 & (15)\\
\hline
Helioseismology & 1996 & 14.16 & -1.63 & -2.52 & 13.62 & -0.662 & -0.120 & (16)\\
Helioseismology & Apr 2002 & 14.04 & -1.70 & -2.49 & 13.49 & -0.672 & -0.119 & (17)\\
\hline
\end{tabular}
\tablebib{
(1)~\citet{Dupree1972}; (2)~\citet{Hara2009}; (3)~\citet{Brajsa2004};
(4)~\citet{Wohl2010};
(5)~{this paper};
(6)~\citet{Pulkkinen1998}; (7)~\citet{Balthasar1986}; (8)~\citet{Sudar2014};
(9)~\citet{Brajsa2002}; (10)~\citet{Howard1984}
(11)~\citet{Brajsa1991};
(12)~\citet{Snodgrass1983}; (13)~\citet{Komm1993};
(14)~\citet{Howard1970}; (15)~\citet{Snodgrass1984}; (16)~\citet{Schou1998}; (17)~\citet{Komm2004}
}
\end{table*}
In Table~\ref{Tab_comp_fit} we show a comparison of the solar differential profile
(Eq.~\ref{Eq_difrot_profile}) from a number of different sources, including
the results from this paper (Table~\ref{Tab_coef_fit}). Since we found no
statistically significant difference between Northern and Southern hemisphere
we only include the results for both hemispheres combined.
Results in Table~\ref{Tab_comp_fit} come from a wide variety of different
techniques, tracers and instruments.

\citet{Snodgrass1984} suggested that the rotation profile should be expressed
in terms of Gegenbauer polynomials since they are orthogonal on the disk.
This eliminates the cross-talk problem between coefficients in Eq.~\ref{Eq_difrot_profile}.
Using the expansion in terms of Gegenbauer polynomials
solar rotation profile becomes:
\begin{equation}
\omega_{rot}(b)=A_{G}T_{0}^{1}(\sin b) + B_{G}T_{2}^{1}(\sin b) + C_{G}T_{4}^{1}(\sin b),
\label{Eq_difrot_profile_G}
\end{equation}
where $A_{G}$, $B_{G}$ and $C_{G}$ are coefficients of expansion and $T_{0}^{1}(\sin b)$,
$T_{2}^{1}(\sin b)$ and $T_{4}^{1}(\sin b)$ are Gegenbauer polynomials as
defined by \citet{Snodgrass1985} in their equation (2).

As \citet{Snodgrass1985, Snodgrass1990} pointed out, the relationship between coefficients
$A$, $B$ and $C$ from standard rotation profile (Eq.~\ref{Eq_difrot_profile})
and coefficients $A_{G}$, $B_{G}$ and $C_{G}$ from Eq.~\ref{Eq_difrot_profile_G}
is linear. Therefore, it is not necessary to recalculate the fits using
Gegenbauer polynomials, we can compute $A_{G}$, $B_{G}$ and $C_{G}$
directly from $A$, $B$ and $C$.
We used the relationship given in \citet{Snodgrass1985} (their
equation (4)) since there seems to be a typo for a similar relationship
for coefficients $C$ and $C_{G}$ in \citet{Snodgrass1990}.
In Table~\ref{Tab_comp_fit} we also show the values of coefficients
$A_{G}$, $B_{G}$ and $C_{G}$.

Our rotational profile results are roughly consistent with 
all the previously published work we surveyed (Table~\ref{Tab_comp_fit}). 
The accuracy of our coefficients is lower when compared with other
results, a consequence of our fairly small number of data points ($n=906$).
\citet{Wohl2010}, for example, had more than 50000 data points spanning a 
time interval of 8 years. 
We have used data spanning only 2 days. It is therefore reasonable 
to expect that with
AIA/SDO CBP data we could reach 50000 data points with only 4 months of data 
and achieve similar accuracy in solar rotation profile coefficients. This means
that with AIA/SDO data it should be possible to measure
rotation profile several times per year and track possible changes in
solar surface differential rotation directly with a very simple tracer
method. This is also true for meridional motion and Reynolds stress, both
of which probably vary over the solar cycle \citep[cf. e.g][]{Sudar2014}.

\section{Summary and Conclusion}

Using 19.3nm data from the SDO/AIA instrument at 10 min cadence we have
identified a large number of CBPs, resulting in 906 rotation velocity
measurements. We obtained a fairly good differential solar rotation
profile in spite of the fact that we used data spanning only two days.
The large density of data points in time
is a result of several factors. The
instrument itself (SDO/AIA) has better spatial resolution and is
capable of high cadence ($<$5 min). For comparison, 
SOHO/EIT 28.4 nm channel
had a cadence of two images every 6 hours \citep{Wohl2010}. In this work,
the high cadence enabled us to track
and measure velocities of short lived CBPs which couldn't be detected
or accurately tracked
by the comparatively large time interval between successive images in
the SOHO/EIT 28.4 nm channel.
Coupled with the fact that short lived CBPs are more numerous than long
lived ones \citep{Brajsa2008}, this resulted in a very high density of data points in time.
High data density necessitated the use of an automatic procedure to detect
and track CBPs. The segmentation algorithm used here proved to be completely adequate
for the task, as similar algorithms have elsewhere
\citep{McIntosh2005,McIntosh2014b,McIntosh2014a}.

The surface rotation profile and its accuracy obtained by helioseismology is seldom given
in a form suitable for comparison with those obtained by
tracer measurements. We can estimate from the number of published significant digits in
the results \citet{Schou1998, Komm2004}, given in Table~\ref{Tab_comp_fit}, that the accuracy
is of the same order as better quality tracer measurements. \citet{Zaatri2009} published the error
for the coefficient $B$ (see Eq.~\ref{Eq_difrot_profile}) and the value of 0.01 from that paper
is in agreement with our estimate above.

It is quite conceivable that errors in the differential rotation
profile coefficients would drop significantly when more data is used.
From our analysis, we can expect to obtain 400-500 velocity measurements per day from CBPs using SDO/AIA.
A time interval of 4 months seems adequate to obtain 50000 velocity
measurements, which should be sufficient to match the most accurate
results obtained by tracer methods (for example \citet{Wohl2010}).

CBPs are also very good tracers since they extend to much higher
latitudes than sunspots. They are also quite numerous in all
phases of the solar cycle while sunspots are often absent in the
minimum of the cycle.

This opens up an intriguing possibility of measuring the solar rotation
profile almost from one month to the next over an entire cycle. Such studies could provide
new insight into mechanisms responsible for solar rotation.
We already know that meridional motion exhibits some changes
during the course of the solar cycle, and the same is probably
true for Reynolds stress. \citet{Sudar2014} found by averaging
almost 150 years of sunspot data that meridional motion changes slightly
over the solar cycle and hinted that the Reynolds stresses
are probably changing too.
Here we have found a small asymmetry in rotation profile for two solar
hemispheres and suggested that this might be related to different solar
activity levels in the two hemispheres. This needs to be verified with a larger
dataset though, as the difference in rotation profiles was of low statistical
significance.

The planned SDO mission duration of 5--10
years will cover a large portion of the solar cycle which should result in
enormous amount of velocity data to assist in the understanding of the
nature and variation of solar rotation profile.
Having more detailed temporal resolution and direct results (without
the need to average many solar cycles) could prove to
be very informative.

A time interval of 10 minutes between successive images also offers a
good opportunity to study the evolution of CBPs and possible effect
this might have on the detected surface velocity fields. For example,
\citet{Vrsnak2003} reported that longer lasting CBPs show different
results than short-lived CBPs.

Based on the promising results here, we will use larger datasets
to further exploit the 
potential of SDO/AIA CBP data to determine
meridional motions, rotation velocity residuals, Reynolds stresses
and proper motions in subsequent papers.

\begin{acknowledgements}
This work has received funding from the European Commission FP7
project eHEROES (284461, 2012-2015) and SOLARNET project (312495, 2013-2017)
which is an Integrated Infrastructure Initiative (I3)
supported by FP7 Capacities Programme. It was also supported by Croatian Science
Foundation under the project 6212 Solar and
Stellar Variability.
SS was supported by NASA Grant NNX09AB03G to the Smithsonian
Astrophysical Observatory and contract SP02H1701R from Lockheed-Martin
to SAO.
We would like to thank SDO/AIA science teams for providing the
observations.
We would also like to thank
Veronique Delouille and Alexander Engell for valuable help in preparation of this work.

\end{acknowledgements}

\bibliographystyle{aa}
\bibliography{Rotation} 

\end{document}